\documentclass[sn-mathphys-num]{sn-jnl}

\usepackage{graphicx}%
\usepackage{multirow}%
\usepackage{amsmath,amssymb,amsfonts}%
\usepackage{amsthm}%
\usepackage{mathrsfs}%
\usepackage[title]{appendix}%
\usepackage{xcolor}%
\usepackage{textcomp}%
\usepackage{manyfoot}%
\usepackage{booktabs}%
\usepackage{algorithm}%
\usepackage{algorithmicx}%
\usepackage{algpseudocode}%
\usepackage{listings}%
\usepackage{hyperref}%

\theoremstyle{thmstyleone}%
\theoremstyle{thmstyletwo}%

\theoremstyle{thmstylethree}%

\newcommand{\be}{\begin{equation}}
\newcommand{\ee}{\end{equation}}
\newcommand{\D}{\mathcal{D}}
\newcommand{\bea}{\begin{align}}
\newcommand{\eea}{\end{align}}

\begin{document}

\title{Evaluating Feynman Integrals through differential equations and series expansions}


\author*[1,2]{\fnm{Tommaso} \sur{Armadillo}}\email{tommaso.armadillo@uclouvain.be}

\affil[1]{\orgdiv{Centre for Cosmology, Particle Physics and Phenomenology (CP3)}, \orgname{Universit\`e catholique de Louvain}, \orgaddress{\street{Chemin du Cyclotron, 2}, \city{Louvain-la-Neuve}, \postcode{B-1348}, \country{Belgium}}}

\affil[2]{\orgdiv{Dipartimento di Fisica “Aldo Pontremoli”}, \orgname{University of Milano and INFN, Sezione di Milano}, \orgaddress{\city{Milano}, \postcode{I-20133}, \country{Italy}}}


\abstract{
We review the method of the differential equations for the evaluation of multi-loop Feynman integrals. In particular, we focus on the series expansion approach for solving the system of differential equation and we discuss how to perform the analytical continuation of the result to entire (complex) phase-space. This approach allow us to consider arbitrary internal complex masses.
This review is based on a lecture given by the author at the 'Advanced School \& Workshop on Multiloop Scattering Amplitudes' held in NISER, Bhubaneswar (India) in January 2024.
}




\maketitle

\section{Introduction}\label{sec:introduction}

The Standard Model (SM) is the theory that classifies all elementary particles and describes their interactions. During the years it has proven itself very
successful in explaining and predicting with extreme precisions a big variety of phenomena, spanning several orders of magnitude. The discovery of the existence of the Higgs boson~\cite{CMS:2012qbp,ATLAS:2012yve}, which has been confirmed in 2012 at the Large Hadron Collider (LHC), is one of its greater success. Despite its incredible achievements, however, the SM has some unsolved problems. For example, it does not include gravity in its description of reality, it is not able to explain dark matter and dark energy nor the matter-antimatter asymmetry. The presence of these problems hints that the Standard Model is not yet complete and an extension is needed. Nowadays there exist multiple theories which are suitable candidates for an extension of the SM, namely Grand Unified Theories (GUT), Supersymmetry, String Theory etc., but no clear evidence of physics beyond the SM has been found. 
Although during the high luminosity run at the LHC we may still be able to make new exciting discoveries, physicists have started to pursue another path to explore the unknown territory of new physics, that is looking for small deviations between theoretical predictions and experimental data. The presence of these small deviations could give, indeed, precious insight on how to extend the SM. For this reason theoretical physicists in the latest years have focused their efforts on obtaining predictions as accurate as possible. 
These calculations, however, are very far from trivial and in order to achieve a percent, or even sub-percent, level of accuracy one must have control on all the sources of theory uncertainties. These include, for instance, Parton Distributions Functions (PDFs), higher order corrections, computed either in Quantum Chromodynamics (QCD) or in the Electro-Weak (EW) sector, and input scheme. The EW sector of the Standard Model, indeed, depends on three independent parameters selected from precisely measured quantities; this choice affects the organization of radiative corrections, and controlling it is crucial for high-precision predictions\footnote{See for example Sec. 5.1 in \cite{Denner:2019vbn}.}.

One of the main bottleneck for the inclusion of higher order corrections is the calculation of the so-called Feynman integrals~\cite{Smirnov:2004ym,Weinzierl:2006qs}. Indeed, in state-of-the-art problems their number can grow up to thousands or hundred of thousands, and hence a general and algorithmic approach is necessary for their evaluation. Fortunately, not all of them are independent one from the other, and by exploiting integration-by-parts (IBPs) identities~\cite{Chetyrkin:1981qh} one is able to find a set of independent integrals, such that all the others can be expressed in terms of these ones. These integrals are called Master Integrals (MIs).
One of the most effective computational tools for evaluating the MIs is the method of differential equations, firstly proposed by Kotikov~\cite{Kotikov:1990kg,Kotikov:1990zs,Kotikov:1990zk} and later improved by Remiddi and Gehrmann~\cite{Remiddi:1997ny,Gehrmann:1999as,Gehrmann:2000zt,Gehrmann:2001ck}. The idea behind this method is that these integrals are functions of kinematics variables and internal masses, and by differentiating with respect to one of those variables one is able to obtain a first order linear differential equation whose unknowns are the MIs we are interested in. The problem, hence, shifts from integrating to solving a differential equation. The solution to these equations can be written in terms of one-dimensional iterated integrals, which, in many cases, correspond to some known classes of functions such as Harmonic Polylogarithms (HPLs)~\cite{Remiddi:1999ew,Gehrmann:2001pz,Maitre:2005uu,Maitre:2007kp,Bonciani:2010ms}. However, when increasing the number of loops, external legs or internal masses, an analytical solution in terms of known classes of functions can become extremely difficult to obtain. For this reason, in latest years the series expansion approach is gaining popularity, thanks also to many public implementations: AMFLow~\cite{Liu:2022chg}, DiffExp~\cite{Hidding:2020ytt}, Line~\cite{Prisco:2025wqs} and SeaSyde~\cite{Armadillo:2022ugh}. Its main idea is to look for a series solution to the differential equations, so that the we are able to evaluate it easily at any point of the phase-space.
A general difficulty that arises when calculating corrections at fixed perturbative order is the presence of internal unstable particles such as Ws and Zs. A complete description of resonances in perturbation theory requires a Dyson summation of self-energy insertions. In particular, it has been observed that the pole of the resummed propagator is a gauge-invariant complex quantity. Therefore it is possible to define the renormalized mass of unstable particles as the complex pole of the resummed propagator. This scheme is called complex mass scheme (CMS)~\cite{Denner:1999gp,Denner:2005fg}, and is necessary to correctly describe the kinematic region near the resonance. In order to perform a complete calculation in the CMS we need to be able to evaluate the master integrals also with complex internal masses, and the series expansion approach to the differential equation method can be exploited to this end.

\section{Feynman Integrals}\label{sec:feynmanintegrals}

The object that we are interested in computing are Feynman Integrals (FIs), i.e. dimensionally regularized integrals~\cite{tHooft:1973mfk} of the following form:
\be
    I_{\boldsymbol{\alpha}}(s_j;d)=\int\prod_{k=1}^l \frac{d^d q_k}{(2 \pi)^{d}} \frac{1}{\D_1^{\alpha_1}\dots\D_n^{\alpha_n}}, 
    \label{eq:fi}
\ee
where $l$ is the number of loops, $d=4-2\epsilon$ is the number of dimensions, $s_j$ are the kinematic invariants that the FIs depend on, $q_k$ are the loop momenta and $\D_i=p_i^2-m_i^2$ are inverse propagators. In particular, $p_i$ is a linear combination of external and loop momenta while $m_i$ is the mass of the particle running in the i-th propagator. Finally, each denominator is raised to an integer power $\alpha_i$, which might also take negative values.

A given set of denominators $\{\D_i\}$ constitutes an integral family and within an integral family, each FI is uniquely identified by the set $\boldsymbol{\alpha}=\{\alpha_1,\dots,\alpha_n\}$ of powers to which each denominator is raised. In the following, we will see that some scalar products might appear in the numerator of the integrand function. In order to bring again the integral in the form of Eq.~\ref{eq:fi}, we have to be able to express each scalar product as a linear combination of denominators and kinematic invariants. Since we can have $l^2+l(n-1)$ independent scalar products, where $n$ is the number of external legs, we need to have the same number of denominators. This means that we might have to define extra denominators, which are called auxiliary denominators. This is better understood with an example.

\subsection{Example: two-loop massless triangle}
\label{sec:examplescalarprod}
We consider the following two-loop massless triangle:
\begin{figure}[H]
\centering
\includegraphics[width=0.4\textwidth]{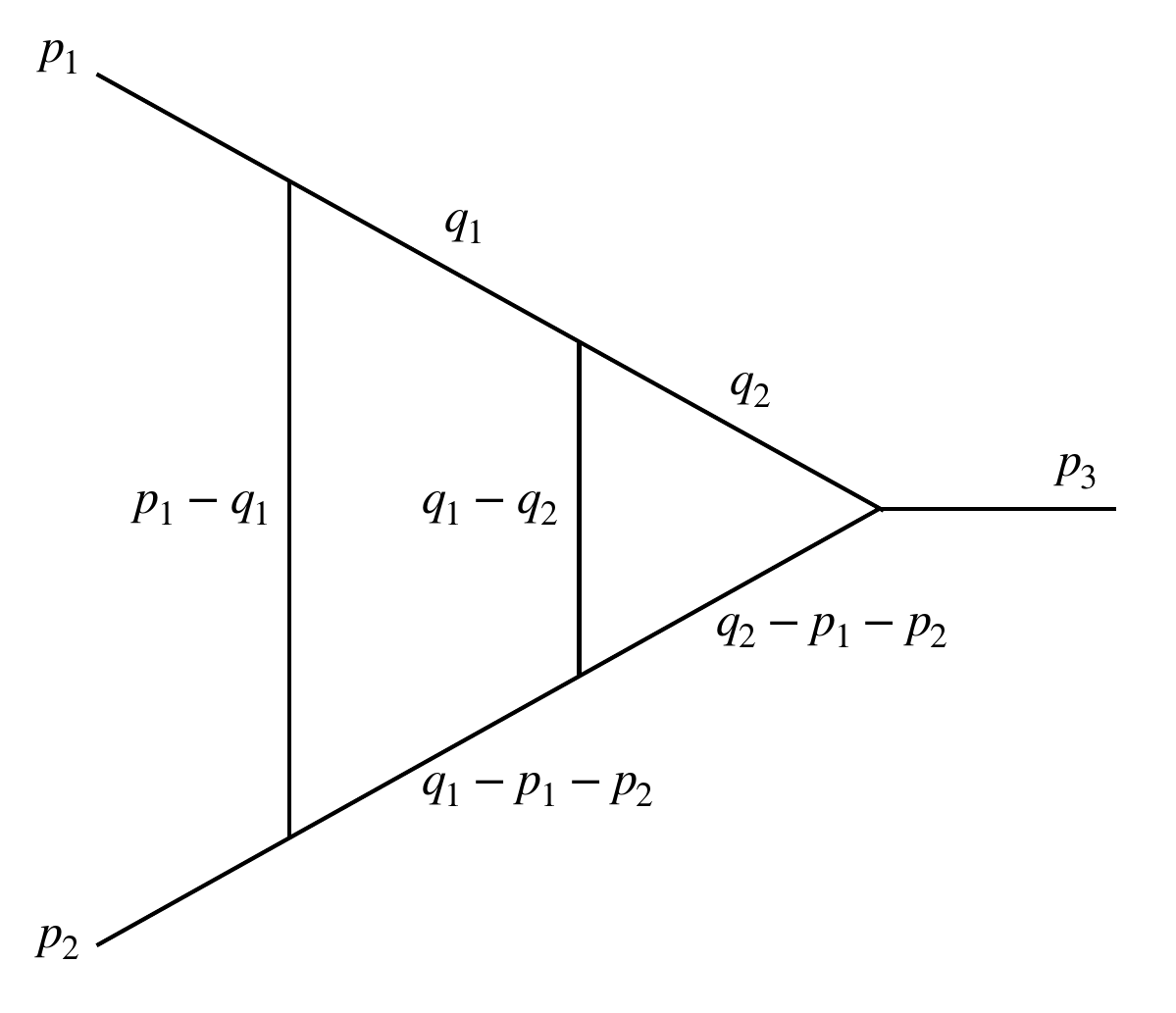}
\end{figure}
\noindent
with $p_1$, $p_2$ and $p_3$ the external momenta satisfying $p_1^2=p_2^2=p_3^2=0$ and $p_1+p_2=p_3$. By labeling $q_1$ and $q_2$ the upper loop propagators, we can fix all the others just by imposing conservation of momentum on each vertex. The list of six denominators then read:
\begin{align}
\nonumber   &\D_1=q_1^2,                &&\D_2=q_2^2,\\ 
\nonumber   &\D_3=(q_1-p_1)^2,          &&\D_4=(q_1-q_2)^2,\\ 
            &\D_5=(q_1-p_1-p_2)^2,      &&\D_6=(q_2-p_1-p_2)^2.
\end{align}
However, we have $7$ possible scalar products. We introduce an auxiliary denominator $\D_7=(q_2-p_1)^2$, so that every scalar product can be written as a combination of $\D_i$. In particular, we find:
\begin{table}[h!]
\centering
\def\arraystretch{1.5}
 \begin{tabular}{|c | c|} 
 \hline
 Scalar Product & Relation \\ 
 \hline\hline
 $q_1^2$ & $\D_1$  \\ \hline
 $q_2^2$ & $\D_2$  \\ \hline
 $q_1\cdot q_2$ & $\frac{1}{2}(\D_1+\D_2-\D_4)$ \\ \hline
 $q_1\cdot p_1$ & $\frac{1}{2}(\D_1-\D_3)$  \\ \hline
 $q_1\cdot p_2$ & $\frac{1}{2}(\D_3-\D_5)$ \\ \hline
 $q_2\cdot p_1$ & $\frac{1}{2}(\D_2-\D_7)$  \\ \hline
 $q_2\cdot p_2$ & $\frac{1}{2}(\D_7-\D_6)$ \\ \hline
 \end{tabular}
 \caption{Relations between scalar products and denominators for the two-loop massless triangle}
\end{table}

If, at any point of the calculation, we find a scalar product in the numerator, we can bring the FI back to the form of Eq.~\ref{eq:fi} using the relations in the table above. For example:
\begin{align}
\nonumber
\int \frac{d^d q_1}{(2\pi)^d} \frac{d^d q_2}{(2\pi)^d}
\frac{q_1\cdot q_2}{\D_1 \D_2 \D_4 \D_5}&=\frac{1}{2}\int \frac{d^d q_1}{(2\pi)^d} \frac{d^d q_2}{(2\pi)^d}
\frac{\D_1+\D_2-\D_4}{\D_1 \D_2 \D_4\D_5}=\\
&=\frac{1}{2}\left(I_{0,1,0,1,1,0,0}+I_{1,0,0,1,1,0,0}-I_{1,1,0,0,1,0,0}\right).
\end{align}
This will be useful when obtaining the differential equations.

\subsection{Integration-by-parts identities}

During state-of-the-art calculations, one usually has to compute up to hundred of thousands of FIs. However, not all of them are independent, but there exist linear relations among them. These relations go under the name of Integration-by-parts Identities (IBPs) and they follow from Gauss's theorem in d-dimensions:
\be
\int\prod_{k=1}^l \frac{d^d q_k}{(2\pi)^d}
\frac{\partial}{\partial q_i^\mu}\left\{v_\mu 
\frac{1}{\D_1^{\alpha_1}\dots\D_n^{\alpha_n}}
\right\}=0,
\label{eq:gauss}
\ee
where $v_\mu$ can be either a loop or an external momentum. The first step is to write down identities from Eq.~\ref{eq:gauss}, for different values of $v_\mu$ and for all the $\boldsymbol{\alpha}$ that appear in our calculation. 
Then we can solve the system by Gauss substitution rule, i.e. considering each equation, expressing one integral in terms of the others and substitute it in the remaining ones. This procedure can be approached in a systematic way by using Laporta algorithm~\cite{Laporta:2000dsw}. This is done by assigning a weight to each integral and by systematically expressing integrals of higher weight in terms of others with lower weight. Laporta algorithm is extremely simple and elegant and it has been implemented in several public codes, such as AIR~\cite{Anastasiou:2004vj}, Blade~\cite{Guan:2024byi}, FIRE~\cite{Smirnov:2008iw, Smirnov:2023yhb}, Kira~\cite{Maierhofer:2017gsa,Klappert:2020nbg}, LiteRed~\cite{Lee:2013mka}, NeatIBP~\cite{Wu:2023upw} and Reduze~\cite{vonManteuffel:2012np}. Its algebraic complexity, however, grows very fast with the number of loops, external legs and internal masses. 
For this reason, for state-of-the-art problem, Laporta algorithm is used in combination with a more advanced technique, the method of finite fields. The idea behind this method is to to solve the system of IBPs numerically multiple times over finite fields and then, at the end, reconstruct the symbolic rational coefficients for the identities of interest by combining the samples together. The advantage of this approach is that the implementation of modular arithmetic in statically typed languages such as C or C++ is fast, exact and suitable for parallelization since each sample is independent from the others. The idea was originally introduced in~\cite{vonManteuffel:2014ixa}, and later implemented in public packages such as FiniteFlow~\cite{Peraro:2019svx}, FireFly~\cite{Klappert:2019emp} and Ratracer~\cite{Magerya:2022hvj}.
In general, obtaining compact IBPs in a reasonable amount of time is one of the bottleneck of the calculations.

After the implementation of IBPs we are able to express every integral, belonging to the a given integral family, in terms of a finite number of FIs which are called Master Integrals (MIs). These Master Integrals play the role of a basis in the space of Feynman integrals. Note that the term basis may be misleading, because a priori we do not know if this set of MIs is minimal, that is if there exist other relations between them and so if they are truly independent on from the other, anyway this term gives a good idea of what is going on.

\section{Differential Equations}\label{sec:diffeq}

In this section we will present a general method to evaluate the MIs introduced in the previous section: the method of differential equations, which was proposed in a simplified version by A. Kotikov and later improved by E. Remiddi and T. Gehrmann, who successfully applied this method to some significant examples~\cite{Mastrolia:2002gt,Czachor:2001mv,Caffo:1998du,Kotikov:2000ye}.

\subsection{The method}
\label{sec:photonequation}
This method aims at writing down a system of differential equations with respect to one of the kinematic invariants of the problem, and, subsequently, solve it. 
The first step is to differentiate all the MIs w.r.t. one of the kinematic invariants. This is done by exploiting the chain rule:
\be
\frac{\partial}{\partial s_k} = \sum_{i,j=1\atop{i\le j}}^{n-1} \frac{\partial (p_i\cdot p_j)}{\partial s_k} \frac{\partial}{(p_i\cdot p_j)},
\label{eq:chainrule}
\ee
where $n$ is the number of external momenta. The derivative of a MI w.r.t. the scalar product $(p_i\cdot p_j)$ can be computed with the following formulas:
\begin{align}
    \nonumber
    \frac{\partial I_{\boldsymbol{\alpha}}(\boldsymbol{s};d)}{\partial (p_i\cdot p_j)} &= \sum_k \left[\mathbb{G}^{-1} \right]_{kj} p_k \cdot \frac{\partial I_{\boldsymbol{\alpha}}(\boldsymbol{s};d) }{\partial p_i} = 
    \sum_k \left[\mathbb{G}^{-1} \right]_{ki} p_k \cdot \frac{\partial I_{\boldsymbol{\alpha}}(\boldsymbol{s};d) }{\partial p_j}, \\
    \frac{\partial I_{\boldsymbol{\alpha}}(\boldsymbol{s};d)}{\partial p_i^2} &=\frac{1}{2} \sum_k \left[\mathbb{G}^{-1} \right]_{ki} p_k \cdot \frac{\partial I_{\boldsymbol{\alpha}}(\boldsymbol{s};d) }{\partial p_i},
\end{align}
where $\mathbb{G}=\mathbb{G}(p_1,\dots,p_{n-1})$ is the Gram matrix, i.e.
\be
\mathbb{G}(p_1,\dots,p_{n-1}) =
\begin{pmatrix}
    p_1^2   &  p_1\cdot p_{2} & \dots    &      p_1\cdot p_{n-1}    \\
    p_1\cdot p_{2} & p_2^2 & \dots    &      p_2\cdot p_{n-1} \\
    \vdots  &  & \ddots   &      \vdots              \\
    p_1\cdot p_{n-1} & p_2\cdot p_{n-1}   &   \dots    &      p_{n-1}^2
\end{pmatrix}
\ee
Note that when applying the operator in Eq.~\ref{eq:chainrule} to an integral of the form of Eq.~\ref{eq:fi}, only two things can happen. The power $\alpha_j$ to which one inverse denominator is raised can increase by 1 and/or a scalar product of loop and external momenta can appear in the numerator. The latter is then reduced to a combination of $\mathcal{D}_i$ as shown is Sec.~\ref{sec:examplescalarprod}. During this procedure only denominators belonging to the given integral family can appear, meaning that is possible to express the derivative of a MI in terms of a linear combination of other FIs belonging to the same integral family. 
Finally, we can use again IBPs identities to reduce this combination of FIs to a linear combination of MIs. By repeating this process for all the MIs we end up with a system of homogeneous first order linear differential equations, to which the MIs are a solution.

\subsection{Example: 1L QED vertex}
\label{sec:example1LQED}
Let us consider the following integral family, which appears in the calculation of the 1-loop vertex correction in Quantum Electrodynamics (QED):
\begin{equation}
    I_{\alpha_1,\alpha_2,\alpha_3}(s;d) = \int \frac{d^d q}{ (2\pi)^d} \frac{1}{\left[q^2\right]^{\alpha_1} \left[(q+p_1)^2-m^2\right]^{\alpha_2}\left[(q-p_2)^2-m^2\right]^{\alpha_3}}.
\end{equation}
This integral family has only two MIs, namely $I_{1,0,0}$ and $I_{1,1,1}$, two independent momenta, $p_1$ and $p_2$, and one kinematic invariant $s=(p_1+p_2)^2$. The Gram matrix reads:
\be
\mathbb{G}=
\begin{pmatrix}
    m^2     &   \frac{s-2m^2}{2} \\
    \frac{s-2m^2}{2}    &   m^2
\end{pmatrix}\;
, 
\qquad
\mathbb{G}^{-1}=\frac{1}{s(4m^2-s)}
\begin{pmatrix}
    4m^2     &   4m^2-s \\
    4m^2-s    &   4m^2
\end{pmatrix}.
\ee
The first integral does not depend on $s$, so its differential equation is trivial. For the second one we have:
\be
\frac{\partial I_{1,1,1}}{\partial s} = \frac{1}{2} \frac{\partial I_{1,1,1}}{\partial p_1\cdot p_2} = 
\left[
\frac{(p_1+p_2)^\mu}{s} + \frac{(p_1-p_2)^\mu}{s-4m^2}
\right]
\frac{\partial I_{1,1,1}}{\partial p_1^\mu}.
\label{eq:diffeqI11}
\ee
The derivative w.r.t. $p_1^\mu$ in Eq.~\ref{eq:diffeqI11} reads:
\begin{align}
    \frac{\partial I_{1,1,1}}{\partial p_1^\mu} &= \int \frac{d^d q}{(2\pi)^d} \frac{\partial}{\partial p_1^\mu} \left( \frac{1}{\mathcal{D}_1 \mathcal{D}_2\mathcal{D}_3}\right) = \int \frac{d^d q}{(2\pi)^d} \frac{-2 q^\mu}{\mathcal{D}_1 \mathcal{D}_2^2 \mathcal{D}_3}.
\end{align}
Now we can insert it in Eq.~\ref{eq:diffeqI11}, use $q\cdot p_1=(\mathcal{D}_2-\mathcal{D}_1)/2$ and $q\cdot p_2=(\mathcal{D}_1-\mathcal{D}_3)/2$, and simplify the expression. In the end we obtain:
\be
\frac{\partial I_{1,1,1}}{\partial s} = 
\frac{1}{s-4m^2} I_{0,2,1} + 
\frac{s-2m^2}{s(s-4m^2)} I_{1,1,1}+ 
\frac{2m^2}{s(s-4m^2)} I_{1,2,0}.
\label{eq:diffeqtria}
\ee
Finally, we use again the IBPs in order to reduce the integral in the r.h.s. of Eq.~\ref{eq:diffeqtria} to a combination of the two MIs of the family:
\be
\frac{\partial}{\partial s}
\begin{pmatrix}
    I_{1,0,0}\\
    I_{1,1,1}
\end{pmatrix}
=
\begin{pmatrix}
      0      &    0   \\
      \frac{\epsilon-1}{s m^2 (s-4m^2)}      &    \frac{2m^2-s-s\epsilon}{s (s-4m^2)}
\end{pmatrix}
\begin{pmatrix}
    I_{1,0,0}\\
    I_{1,1,1}
\end{pmatrix}.
\label{eq:matrixeq}
\ee
Before moving on, a few comments are in order. The first one regards the singular structure of the MIs. In particular, we observe that this can be read from the poles of the coefficient matrix in Eq.~\ref{eq:matrixeq}. Indeed, we see that there are two of them, one for $s=0$ and the other for $s=4m^2$. The first one is a pseudo-threshold, while the second is the physical threshold associated to the on-shell production of the two internal fermions.

A second comment regards the number of variables that appear in the system. A common procedure is to introduce adimensional variables, defined as ratios of kinematical invariants. This is done in order to reduce by one the total number of variables in the differential equations, thus simplifying the solution procedure. For example, one could introduce the variable $x=s/m^2$, so that the system in Eq.~\ref{eq:matrixeq} becomes
\be
\frac{\partial}{\partial x}
\begin{pmatrix}
    I_{1,0,0}\\
    I_{1,1,1}
\end{pmatrix}
=
\begin{pmatrix}
      0      &    0   \\
      \frac{\epsilon-1}{x (x-4)}      &    \frac{2-x-x\epsilon}{x (x-4)}
\end{pmatrix}
\begin{pmatrix}
    I_{1,0,0}\\
    I_{1,1,1}
\end{pmatrix}.
\label{eq:matrixeqx}
\ee

\subsection{Boundary Conditions}
Before diving into how to solve the system of differential equations let us address the problem of finding the boundary conditions. 
The boundary conditions, in general, could be given in three different forms. The first one is as the value of MIs in a particular point in the phase-space. In this case we can construct a well-defined Cauchy problem. In order to obtain the boundary conditions in this form we need to solve the MIs explicitly in that particular point of the phase-space, either analytically or numerically. If we choose to tackle the problem analytically, one usually chooses a point in the phase space in which most of the kinematic invariants vanish, so that the integrals are easier to evaluate. However, a Feynman integral usually develops divergences as we approach such a point, and hence by simply plugging these values into the integrand, we would not obtain the correct asymptotic limit. One possible solution to this problem is given by the method of expansion by regions~\cite{Smirnov:1991jn,Beneke:1997zp,Heinrich:2021dbf}. Another possibility is to obtain them numerically in the Euclidean region. This can be done, for example, with pySecDec~\cite{Borowka:2017idc}, which parametrises the integral and then perform the parametric integration with Monte-Carlo techniques. The problem of this approach is that the numerical precision is limited. Finally, one can use the Auxiliary Mass Flow method, implemented in the Mathematica package AMFlow~\cite{Liu:2022chg}. 

There are, then, other two possibilities for fixing the free parameter that appears in the solution. It may happen, indeed, that from independent arguments we know that the integral must be regular at a particular point of the phase-space, usually a pseudo-threshold. In this cases, it might be possible to impose the finiteness of the solution to completely fix the free parameter. A final possibility is to fix the asymptotic behavior of the solution on a threshold. In particular, a general observation we can do about Feynman integrals is that their divergences are no more than logarithmic. In some cases fixing the coefficient of the logarithmic term of the solution might be sufficient to completely determine it. The choice of a method or the other depends on the particular case we are considering.

\subsection{Epsilon expansion}

From now on, we will consider only one kinematic variable at the time. This choice will simplify the calculations and will be made clearer when discussing the analytic continuation of the solution. To this end, we suppose that the boundary conditions are imposed in $\boldsymbol{\tilde s}=\{\tilde s_1,\dots,\tilde s_m\}$ and that we are solving the system with respect to the variable $s_j$. In the system we perform the substitution 
\be
    s_i \to \tilde s_i\; \qquad \text{with}\; i\neq j
\ee
so we obtain a system with only one kinematic variable $s_j$. From this moment on, for simplicity, we will drop the index $j$. So the system, in general, takes the following form:
\be
    \frac{\partial}{\partial s} \Vec{I}(s;\epsilon) = \boldsymbol A (s;\epsilon) \Vec{I}(s;\epsilon),
    \label{eq:system}
\ee
where $\Vec{I}(s;\epsilon)$ denotes the vector of MIs.
First of all, the system needs to be $\epsilon$-expanded. To this end let us write:
\be
    \Vec{I}(s;\epsilon) = \sum_{k=\epsilon_{min}}^\infty \Vec{I}^{(k)}(s) \epsilon^k,
\ee
here we assumed that the expansion starts at order $\epsilon_{min}$. The value of $\epsilon_{min}$ can be determined from the boundary conditions. Along with $\Vec{I}$ we expand also the coefficient matrix $\boldsymbol{A}$.
\be
    \boldsymbol{A}(s;\epsilon) = \sum_{k=0}^\infty \boldsymbol{A}^{(k)}( s) \epsilon^k
\ee
We can always assume that there are no poles in $\epsilon$, since they can be removed by rescaling the basis $\Vec{I}$ by an overall power of $\epsilon$. Using this expansion in the Eq.~\ref{eq:system} and collecting order-by-order the different powers in $\epsilon$ we obtain a tower of equations
\be
    \partial_s \Vec{I}^{(k)}(s) = \boldsymbol{A}^{(0)} (s) \Vec{I}^{(k)}(s) + \sum_{j=\epsilon_{min}}^{k-1}\boldsymbol{A}^{(k-j)} (s) \Vec{I}^{(j)}(s).
    \label{eq:systemexpanded}
\ee
An important thing to notice is that in the equations for $\Vec{I}^{(k)}$ appear $\Vec{I}^{(j)}$ with $j< k$ but not with $j>k$. So we can start from $k=\epsilon_{min}$ and solve the system to obtain $\Vec{I}^{(\epsilon_{min})}$. Now we substitute $\Vec{I}^{(\epsilon_{min})}$ in the equation for $\Vec{I}^{(\epsilon_{min}+1)}$ and solve it with respect to $\Vec{I}^{(\epsilon_{min}+1)}$ and we proceed recursively up to the desired order in $\epsilon$. 

\section{Series Expansion Approach}

In this section we will discuss how to solve the system for each order in $\epsilon$ using the series expansion approach. 
The method was first introduced in Ref.~\cite{Moriello:2019yhu} and was shortly after implemented in the public Mathematica package DiffExp~\cite{Hidding:2020ytt}. The idea of this approach consists in expanding the equations both in $\epsilon$ and in a kinematic variable $s$, so that the problem of solving a differential equation reduces to an algebraic one.
Its main advantage with respect to analytic ones, is that the complexity of the latter grows fast when increasing the number of external legs and/or the number of internal scales. 
While for the series expansion approach, since at each step of the calculation we are dealing with power series and logarithms, all the steps can be carried out analytically and in a systematic way, independently from the mathematical structure of the problem. Moreover, the numerical precision of the solution can be directly controlled by the number of terms that we consider, meaning that, provided that we have infinite time and space, we could achieve, in principle, arbitrary precision. Finally, once we have the solution, it can be evaluated numerically in a negligible amount of time.
Power series, however, have some drawbacks. The biggest one being that series have a limited radius of convergence, meaning that we have to provide an algorithm for performing the analytic continuation of the result. This will be the focus of Sec.~\ref{sec:analyticcont}. 

\subsection{Bottom-to-top approach and solution of equations}

Let us look closer at the system in Eq.~\ref{eq:systemexpanded}, i.e. let us consider a fix order $\epsilon^{\Bar{k}}$ and let us assume that we already solved all the ones for $k<\Bar{k}$. This is a system of first order linear differential equations, in which the equations are in principle coupled. However, in many cases by choosing an appropriate base of MIs, the system in Eq.~\ref{eq:systemexpanded} can be casted in an upper-triangular form, i.e.
\be
    \frac{\partial }{\partial s}
    \begin{pmatrix}
        I_1^{(\Bar{k})} \\
        I_2^{(\Bar{k})} \\
        I_3^{(\Bar{k})} \\
        \vdots\\
        I_n^{(\Bar{k})}
    \end{pmatrix}
    =
    \begin{pmatrix}
        \star   & \star     & \star & \dots & \star \\
        0       & \star     & \star & \dots & \star \\
        0       & 0     & \star & \dots & \star \\
        \vdots & \vdots & \vdots & \ddots & \vdots\\
        0 & 0 & 0 & \dots & \star \\
    \end{pmatrix}
    \begin{pmatrix}
        I_1^{(\Bar{k})} \\
        I_2^{(\Bar{k})} \\
        I_3^{(\Bar{k})} \\
        \vdots\\
        I_n^{(\Bar{k})}
    \end{pmatrix}
    +
    \begin{pmatrix}
        \star \\
        \star \\
        \star \\
        \vdots\\
        \star
    \end{pmatrix}.
    \label{eq:triangle}
\ee
If the system is in the form of Eq.~\ref{eq:triangle}, it can be solved using a bottom-to-top approach. Indeed, the last equation contains only $I^{(\Bar{k})}_n$, so we can solve  for it and substitute the solution into the second to last equation, solve for $I^{(\Bar{k})}_{n-1}$ and proceed recursively until we solve the system completely.

At each step of the process we have to solve a first order linear differential equation, with a given boundary condition.
\be
    \begin{cases}
    \frac{\partial}{\partial s} f(s) = a(s) f(s) + h(s),\\
    f(s_0) = \tilde f
\end{cases}
\label{eq:cauchy}
\ee
where for simplicity we substituted $I_i^{(\Bar{k})} \equiv f$ and $a(s)$ and $h(s)$ are rational functions of $s$. For solving Eq.~\ref{eq:cauchy} we can proceed by solving the homogeneous equation, finding a particular solution and, lastly, by imposing the boundary condition. 

\subsection{Frobenius method}
In order to solve the homogeneous equation we use the Frobenius method. This is a general technique for solving a homogeneous ordinary differential equation
\be
    \frac{\partial}{\partial s} f(s) = a(s) f(s)
    \label{eq:homo}
\ee
around a point $s=s_0$. Without loss of generality from now on we consider $s_0 = 0$.
Let us consider the following ansatz for the solution:
\be
    f_{hom}(s) = s^r \sum_{m=0}^\infty c_m s^m
    \label{eq:ansatz}
\ee
where $c_m$ are complex coefficients and $r$ is a rational exponent. The term $s^r$ has the role to take into account for some divergent terms in the solution. We series expand the function $a(s)$ in the differential equations, then by substituting Eq.~\ref{eq:ansatz} into Eq.~\ref{eq:homo}, and collecting the different terms based on different powers of $s$, we obtain a set of algebraic equations for the coefficients $c_m$. At leading order in $s$, the equation is a non-trivial polynomial equation for $r$, which is called indicial equation. After fixing $r$ we may (recursively) solve for all $c_m$ with $m \ge 1$. In the end there is one last free parameter, namely $c_0$. An example of application of the Frobenius method is given in \ref{sec:example}.

\subsection{Variation of parameters}
Once we have the solution for the homogeneous equation, we can obtain a particular solution using the variation of parameters method. We consider the following ansatz for a particular solution:
\be
f_{part}(s) = C(s) f_{hom}(s)
\label{eq:ansatzpart}
\ee
where $C(s)$ is a function to be determined. Let us substitute Eq.~\ref{eq:ansatzpart} into Eq.~\ref{eq:cauchy}
\be
C'(s) f_{hom}(s) + C(s) f_{hom}'(s) = a(s) C(s) f_{hom}(s) + h(s)
\ee
but from Eq.~\ref{eq:homo} we see that $f_{hom}'(s) =  a(s) f_{hom}(s)$ hence we are left with:
\be
C'(s) f_{hom}(s) = h(s).
\ee
Finally we invert it to obtain $f_{part}(s)$:
\be
f_{part}(s) = f_{hom}(s) \int_{\tilde s}^{s} h(s') f_{hom}^{-1}(s') ds'.
\label{eq:solpart}
\ee
Now that we have a particular solution we can construct the general one by summing the homogeneous together with the particular one:
\be
f(s) = c f_{hom}(s) + f_{part}(s)
\ee
where $c$ is a complex constant that is determined imposing the boundary condition.
Note that since $f_{hom}$ is a series, after expanding $h(s)$, the product $h f_{hom}^{-1}$ is still a series and hence it can be integrated analytically. 

\subsection{Example: 1L QED Vertex}
\label{sec:example}
Let us illustrate how this works in practice by considering the order $1/\epsilon$ of the system for the 1-loop QED Vertex, given in Eq.~\ref{eq:matrixeqx}:
\be
\begin{cases}
&\frac{\partial}{\partial x} B_1^{(-1)}=0\\
&\frac{\partial}{\partial x} B_2^{(-1)}=
-\frac{1}{x\;(x-4)}B_1^{(-1)} - \frac{x-2}{x\;(x-4)}B_2^{(-1)}\\
&B_1^{(-1)}(0)=1\\
&B_2^{(-1)}(0)=1/2
\end{cases}
\ee
where for readability we defined $B_1\equiv I_{1,0,0}$ and $B_2\equiv I_{1,1,1}$. The system is indeed in a triangular form. In particular, the first equation is trivial and gives $B_1^{(-1)}=1$. We start solving the second one by considering the ansatz
\be
B_2^{(-1),hom}(x)=x^r\sum_{i=0}^\infty c_i x^i
\ee
Substituting in the homogeneous equation and collecting different powers of $x$ we find:
\be
\begin{cases}
&r\;c_0=-\frac 1 2 c_0\\
&(1+r)\;c_1=\frac 1 8 - \frac{c_1}{2}\\
&(2+r)\;c_2=\frac{1}{32} \left(1 + 4 c_1 - 16 c_2\right)\\
&(3+r)\;c_3=\frac{1}{128} \left(1 + 4 c_1 + 16 c_2 - 64 c_3\right)\\
&\dots
\end{cases}
\ee
The indicial equation gives $r=-1/2$ and then by recursively solving for all the $c_i$ we find:
\be
B_2^{(-1),hom}(x)=\frac{c_0}{\sqrt{x}}\left( 1 + \frac{1}{8} x -\frac{3}{128} x^2 +\frac{5}{1024} x^3 + \mathcal{O}(x^{4})\right).
\ee
In order to find the particular solution we use Eq.~\ref{eq:solpart}:
\be
    B_2^{(-1),part}(x)=\frac{1}2 + \frac x {12} + \frac{x^2}{60} + \frac{x^3}{280}+\mathcal{O}(x^4)
\ee
The complete solution is hence:
\begin{align}
&B_2^{(-1)}(x)=B_2^{(-1),hom}(x)+B_2^{(-1),part}(x) = \\
\nonumber&\quad=c_0 x^{-1/2} \left(1 + \frac{x}{8} + \frac{3 x^2}{128} + \frac{5 x^3}{1024}+\mathcal{O}(x^4)\right)+
\left(\frac{1}2 + \frac x {12} + \frac{x^2}{60} + \frac{x^3}{280}+\mathcal{O}(x^4)\right)
\end{align}
Finally, we fix the boundary condition $c_0=0$. 

Note that, in this case knowing that the solution was regular for $x=0$ would have been sufficient for fixing the boundary condition. 
From the coefficient matrix we read that there are two possible singular points, i.e. $x=0$ and $x=4$. We just found out that the solution for $x=0$ is regular, meaning that $x=0$ is a pseudo-threshold. We could have also chosen to solve the system around a non problematic point, like for example $x=1$. In this case, we are guaranteed that the solution is always a simple Taylor series. Lastly, we could have solved the equation on top of a physical threshold, like $x=4$. In this case the final solution would have contained terms like
\be
\frac{1}{x-4}\;, \qquad \log(x-4)
\ee
which can arise from the integration in Eq.~\ref{eq:solpart} or, at higher orders in $\epsilon$, from the non-homogeneous term in the equation. 

\begin{figure}
    \centering
    \includegraphics[width=0.8\linewidth]{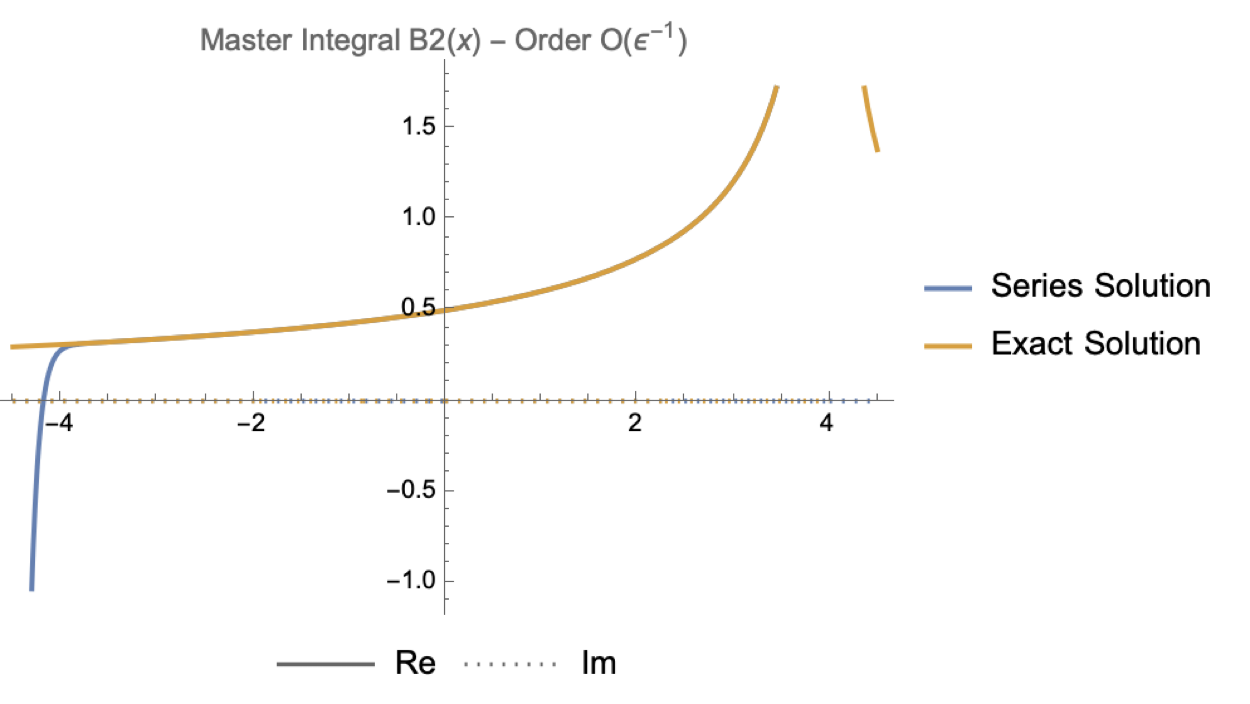}
    \caption{Comparison of the series solution for $B_2^{(-1)}$ against the exact solution. The series is centered in $x=0$ and the closest singularity is $x=4$. This implies that the series converges in the interval $(-4,4)$, which can be seen from the rapidly growing of the solution around $x=-4$.}
    \label{fig:series1LQED}
\end{figure}

Finally a comment regarding the convergence of the series centered in $x=0$. In particular, the radius of convergence is given by the distance between the center of the series with the closest singularity. In this case the radius of convergence is $4$, which means that the series converges in the interval $(-4,4)$. This can be seen explicitly in Fig.~\ref{fig:series1LQED}.

\subsection{Coupled Equations}
In some practical case might be too complicated, or even not possible if the problem has an elliptic nature, to cast the system of differential equations in the form of Eq.~\ref{eq:triangle}. In this section we will discuss how to generalize the methods introduced in the previous sections to a system of coupled equations.
Let us start by considering the following system of coupled equations
\be
\begin{cases}
&B_1'(x)=  \frac{B_1(x)}{x} - \frac{3 B_2(x)}{x} +\left(\frac{1}{2} - \frac{9}{x}\right)\\
&B_2'(x)= - \frac{2 (x-3) B_1(x)}{x (x-9) (x-1)} + \frac{2 (5 x-9) B_2(x)}{  x (x-9) (x-1)}
-\frac{648 + (4 \pi^2-273) x + 27 x^2}{12 x (x-9) (x-1)}\\
&B_1(1)=\frac{59}{8}+\frac{\pi^2}{4}\\
&B_2(1)=\frac{\pi^2}{12}-\frac{1}{2}
\end{cases},
\ee
and we try to solve it around $x=1$. In order to do so we consider the following ansatz for the solution of the homogeneous equation:
\begin{align}
\nonumber&B_1^{hom}(x)= (x-1)^r \; \sum_{i=0}^\infty a_i \;(x-1)^i\\
&B_2^{hom}(x)= (x-1)^r \; \sum_{i=0}^\infty b_i \;(x-1)^i.
\label{eq:ansatzsystem}
\end{align}
Now we proceed in the same way as the single equation case, that is we substitute it in the homogeneous equation, we expand everything around $x=1$, collect the different powers of $(x-1)$ and, finally, solve recursively the infinite set of algebraic equations for $a_i$ and $b_i$. 
The result is:
\begin{align}
\nonumber& B_1^{hom}(x)=a_1\left((x-1)^2 - \frac{5 (x-1)^3}{4} + \frac{ 87 (x-1)^4 }{64}+\mathcal{O}(x-1)^5\right)\\
& B_2^{hom}(x)=a_1\left(- \frac{2 (x-1) }{3} + \frac{11 (x-1)^2 }{12} 
- \frac{47 (x-1)^3 }{48} + \frac{97 (x-1)^4}{96}+\mathcal{O}(x-1)^5\right).
\label{eq:homogsolutionsystem}
\end{align}
By looking closely to the solution in Eq.~\ref{eq:homogsolutionsystem} we see that it depends only on one parameter, namely $a_1$. However, this is a system of two differential equations and, as such, we expect two linearly independent solutions. This is related to the fact that the ansatz we chose was not general enough. We replace, hence, the ansatz in Eq.~\ref{eq:ansatzsystem} with a more general one:
\begin{align}
\nonumber&B_1^{hom}(x)= (x-1)^r \; \sum_{i=0}^\infty a_i \;(x-1)^i + \log(x-1)\;(x-1)^r \; \sum_{i=0}^\infty c_i \;(x-1)^i\\
&B_2^{hom}(x)= (x-1)^r \; \sum_{i=0}^\infty b_i \;(x-1)^i + \log(x-1)\;(x-1)^r \; \sum_{i=0}^\infty d_i \;(x-1)^i
\end{align}
which contains also logarithmic terms. By proceeding always in the same way we find:
\begin{align}
\nonumber B_1^{hom}(x)=&
a_0\bigg[1 - \frac{x-1}{2} + \frac{9 (x-1)^3}{128} +\mathcal{O}(x-1)^4 + \\
\nonumber &\qquad+\left(\frac{3 (x-1)^2}{16} - \frac{15 (x-1)^3}{64} +\mathcal{O}(x-1)^4\right) \log(x-1)
\bigg] +\\
\nonumber&\qquad+ a_2\left((x-1)^2 - \frac{5 (x-1)^3}{4} + +\mathcal{O}(x-1)^4\right);\\
\nonumber B_2^{hom}(x)=&a_0\bigg[\frac{1}{2} - \frac{x-1}{16} - \frac{7 (x-1)^2}{128} + \frac{71 (x-1)^3}{1024} +\mathcal{O}(x-1)^4 + \\
\nonumber&\qquad+\left(-\frac{x-1}{8} + \frac{11 (x-1)^2}{64} - \frac{47 (x-1)^3}{256} +\mathcal{O}(x-1)^4\right) \log(x-1)\bigg]+\\
&\qquad +a_2\left(-\frac{2 (x-1)}{3} + \frac{11 (x-1)^2}{12} - \frac{47 (x-1)^3}{48} +\mathcal{O}(x-1)^4\right)
\end{align}
Now we see that the solution correctly depends on two different parameters, namely $a_0$ and $a_2$.
We can, hence, generalize to a system of $n$ homogeneous differential equations:
\be
\Vec{B}^{hom}(x)=(x-x_0)^r \sum_{i=0}^\infty \Vec{c}_{i,0} (x-x_0)^i
+ (x-x_0)^r \sum_{j=0}^m \log^j(x-x_0)\sum_{i=0}^\infty \Vec{c}_{i,j} (x-x_0)^i.
\ee
Firstly, we try to solve the system for $m=0$. If we obtain $n$ linearly independent solutions we are done, otherwise we increase the value of $m$, eventually up to $n-1$. 

\subsection{Variation of parameters for systems}
The generalization of the variation of parameters technique to systems of differential equations is straightforward. Firstly, we organize the solution of the homogeneous equation in a matrix form:
\be
\Vec{B}^{hom}(x)=\mathbf{M}(x)\Vec{a}
\ee
where $\mathbf{M}_{ij}(x)$ is the solution for the $i$-th MI where we put all the free parameters to 0, except for the $j$-th. And $\Vec{a}$ is the vector of free parameters. Then a particular solution is given by:
\be
\Vec{B}^{part}(x)=\mathbf{M}(x)\int_{x_0}^x\mathbf{M}^{-1}(x') \Vec{g}(x') dx'
\label{eq:particular}
\ee
where $\Vec{g}(x')$ is the vector of non homogeneous terms of the equation. By following this procedure for the example introduced in the previous section we obtain:
\begin{align}
\nonumber B_1(x)=&
\frac{59}{8} + \frac{\pi^2}{4}+\frac{3}{8} (x-1) + \frac{1}{4} (x-1)^2 - \frac{1}{3} (x-1)^3 +\mathcal{O}(x-1)^4\\
B_2(x)=&\frac{1}{2} \left(\frac{\pi^2}6-1\right)+\frac{1}{4} (x-1)^2 - \frac{25}{96} (x-1)^3 +\mathcal{O}(x-1)^4.
\end{align}
A few comments are in order. The first one is that within the integration in Eq.~\ref{eq:particular} only terms like $(x-x_0)^p\log^q(x-x_0)$ can appear. In practice, this integration can be performed with the implementation of substitution rules in order to make it faster. A second comment regards the inversion of the matrix $\mathbf{M}^{-1}(x)$. Inverting a matrix is an operation whose computational cost grows cubically with its dimension. For this reason, even if it might not be possible to write down a system in triangular form, it might be worth to cast it in a block triangular form.

\section{Analytic continuation}\label{sec:analyticcont}
Series have a limited radius of convergence, which is given by the distance of its center to the closest singularity. In order to extend the solution outside the region of convergence, we need to provide an algorithm for performing the analytic continuation. 

\subsection{Complex Mass Scheme}
When working with intermediate unstable particles such as $W$ and $Z$ we have to employ a gauge invariant definition for the mass. Such a definition is given by the complex mass scheme, which identifies the mass of the particle through the position of the pole of the propagator and is a well defined and gauge invariant quantity. In practice, for a gauge boson $V$ we consider a complex mass
\be
\mu_V^2=m_V^2-i \Gamma_V m_V
\ee
where $m_V$ is the real mass of the boson and $\Gamma_V$ its width. When working with adimensional variables, like the ones introduced in Sec.~\ref{sec:example1LQED}, they become complex-valued. For this reason, the analytic continuation is discussed in the complex kinematical plane.

\subsection{Singularities and branch cuts}
\begin{figure}
    \centering
    \includegraphics[width=0.4\linewidth]{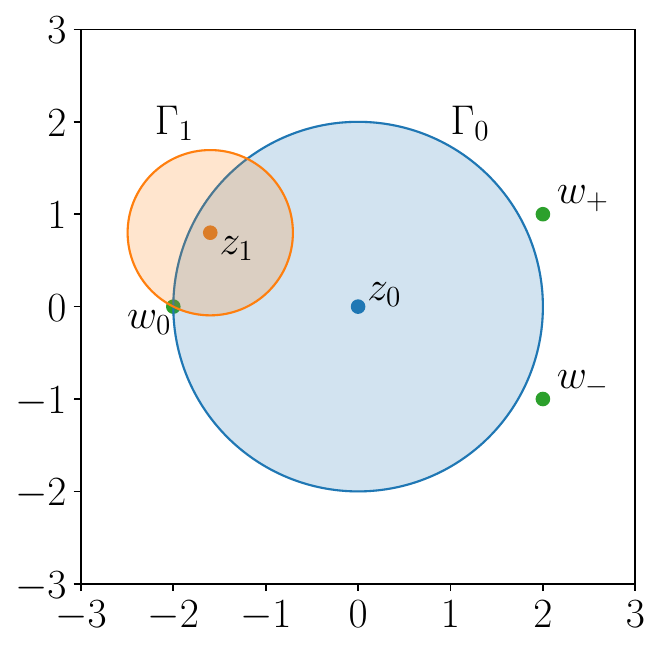}
    \caption{Radius of convergence and example analytic continuation.}
    \label{fig:convergence}
\end{figure}

As already mentioned in previous sections, series have a limited radius of convergence. The latter is given by the distance between the center of the series and closest singular point. This can be seen graphically in Fig.~\ref{fig:convergence}. In particular, we are considering a complex variable $z$ and we are discussing the analytic continuation in the $z$ complex plane. The series is centered in $z_0$ and the radius of convergence is $\rho=\min_{w\in \mathcal{W}}|z-w|$, with $\mathcal{W}=\{w_0,w_+,w_-\}$ the set of all singularities. In the example shown in Fig.~\ref{fig:convergence} the closest singularity is $w_0$, hence, the series converges in the $\Gamma_0$ region.
For extending the solution outside the blue region we have to perform an analytic continuation. This is done by evaluating the solution in a point within $\Gamma_0$, for example $z_1$, and then by solving again the system, this time centered around $z_1$, and using the value we have just computed as a new boundary condition. It is possible to show that this procedure is unique.
\begin{figure}[t]
\centering
\includegraphics[width=0.4\textwidth]{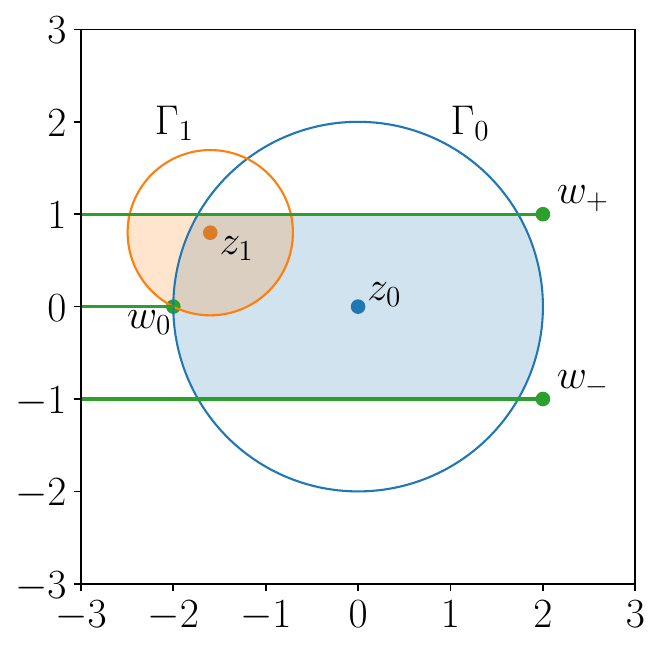}
\includegraphics[width=0.4\textwidth]{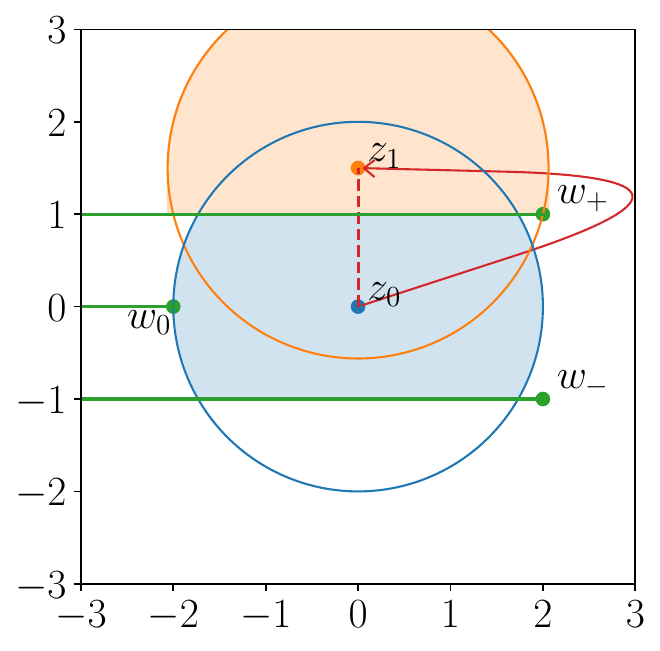}
\caption{\label{fig:cuts}
 Effect of a branch cut on the analytic continuation. It shrinks the area in which the series converges to the desired value (left) and it modifies the path for moving from one point to another (right).}
\end{figure}
We saw also in the previous sections, that the solution might exhibit a logarithmic behavior. This complicates the analytic continuation procedure since it makes the solution a multi-valued function. In order to make it single-valued we have to introduce some branch-cuts, thus specifying which is the physical Riemann sheet on which we evaluate the solution. 
We decided to associate to each logarithmic singularity a branch-cut that starts from the singularity and goes to $-\infty$, parallel to the real axis. While this branch-cut does not modify the region in which the series converges, it shrinks the area in which the solution converges to the desired value, that is the one on the physical Riemann sheet. Looking at left panel of Fig.~\ref{fig:cuts}, while the series still converges in $\Gamma_0$, only in the blue strip it does to the desired value. As a consequence, for extending the solution we have to choose a path that does not cross the branch cut. This is shown in the right panel of Fig.~\ref{fig:cuts}. For reaching $z_1$ we cannot go straight along the dashed line, because we would cross the branch cut and end up on the non physical Riemann sheet. Instead, we should follow the path of the solid red line, that avoids $w_+$ on the right. For readability of the picture, we have not plotted all the intermediate steps. 

\begin{figure}[t]
\centering
\includegraphics[width=0.4\textwidth]{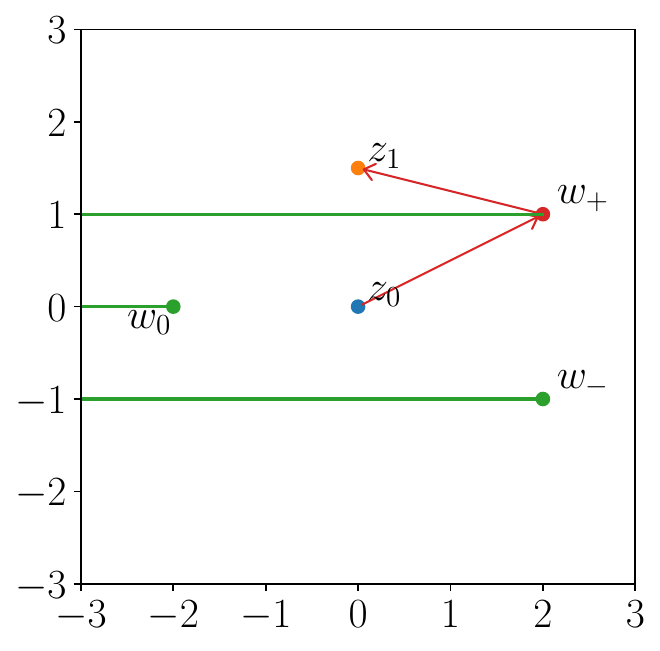}
\includegraphics[width=0.4\textwidth]{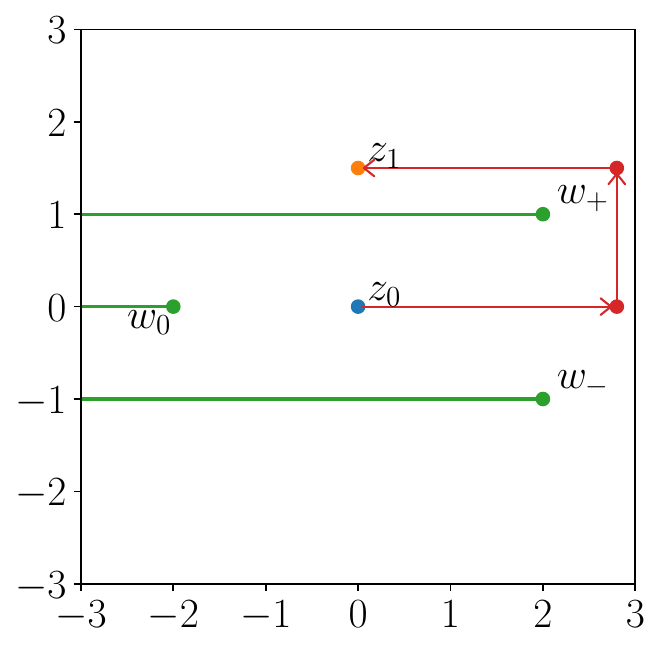}
\caption{\label{fig:path_logVStaylor}
 Example of a Logarithmic expansion on the left, against a Taylor expansion on the right. }
\end{figure}

For following the red line, actually, there are two possibilities. In particular, they differ in the way they treat the singular point, either avoiding it or expanding on top of it. The two strategies are shown in Fig.~\ref{fig:path_logVStaylor} and, while they lead to the same result, they each have pros and cons. If we avoid the singularities (left panel of Fig.~\ref{fig:path_logVStaylor}), at each step, we deal only with Taylor series, hence the solution is simpler and usually quicker to obtain. We dub this strategy Taylor expansion. The second possibility is the logarithmic expansion (right panel of Fig.~\ref{fig:path_logVStaylor}). In this case, the solution might be slower to obtain, however, if the series is centered in $w_+$ the closest singularity is $w_-$, so the radius of convergence is bigger and we can take less intermediate steps. Choosing one strategies over the other is done case by case mainly by looking at the position of the other singularities which are present in the problem.

\begin{figure}[t]
\centering
\includegraphics[width=0.4\textwidth]{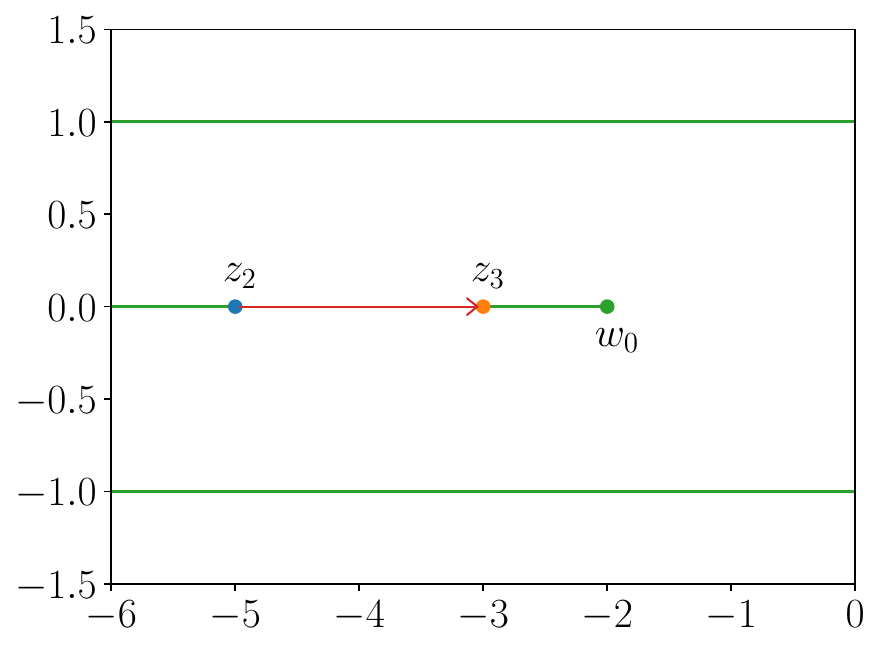}
\includegraphics[width=0.4\textwidth]{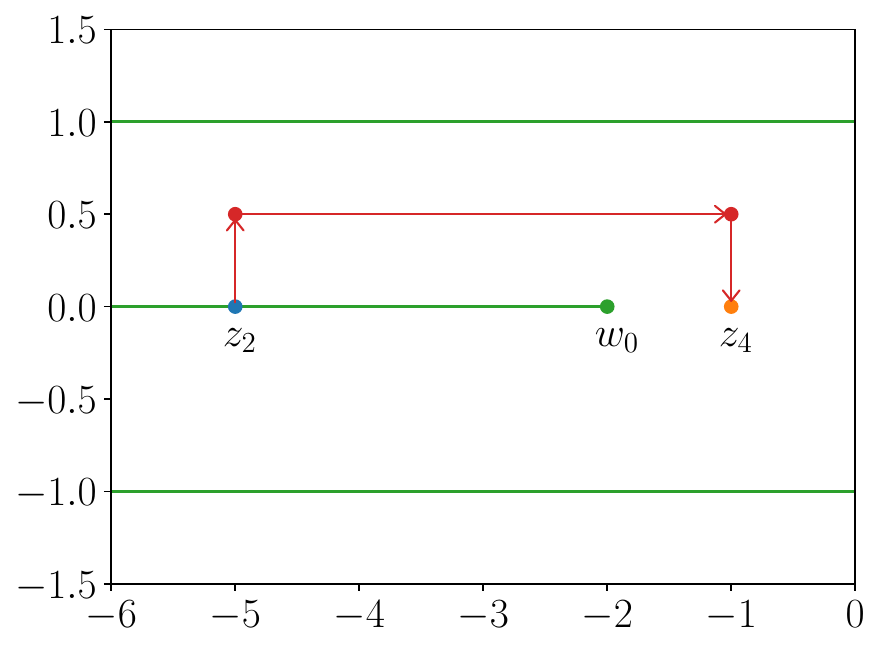}
\caption{\label{fig:realAxis}
Examples of two possible ways to move along a branch cut. In case there is no a singularity between them (left) or there is (right).}
\end{figure}

Finally, we discuss how to move along a branch cut. There are two different cases which are shown in Fig.~\ref{fig:realAxis}. If there is not a singularity between the starting and ending point we can move straight since, by definition, we are never crossing a branch cut. If, on the contrary, we have a singularity between the two points we have to avoid moving either in the upper or lower part of the complex plane. This ambiguity is solved by considering Feynman prescription associated to the kinematical variable. A prescription $+i\delta$ implies an horse-shoe path in the upper half of the complex plane, while a $-i\delta$ in the lower one.

\section{Conclusions}
Multi-loop Feynman integrals are one the main ingredients for the evaluation of higher corrections in high energy physics. One of the most widely used technique to evaluate them in the method of differential equations. However, solving it analytically in terms of known classes of functions can become extremely challenging when increasing the number of loops, external legs and internal scales. For this reason in the last years is gaining popularity the series expansion approach, also thanks to different public implementations. The advantage of this method, with respect to purely analytical techniques, is that it is completely general and easy to automate, meaning that it is blind to the analytical complexity underling the Feynman integrals. In this review we presented how to obtain the system of differential equations and we showed how to solve it using series expansion techniques. Finally, in the last part we reviewed the algorithm for performing the analytic continuation of a series outside its original radius of convergence. 

\bmhead{Acknowledgements}
The work presented has been done in collaboration with Roberto Bonciani, Simone Devoto, Narayan Rana and Alessandro Vicini. T.A. is a Research Fellow of the Fonds de la Recherche Scientifique – FNRS.

\bmhead{Data availability} The package SeaSyde is released under the GNU General Public License and it available for download from https://github.com/TommasoArmadillo/SeaSyde.




\bibliography{sn-bibliography}

\end{document}